\documentclass[fleqn,usenatbib]{mnras}

\usepackage{newtxtext,newtxmath}

\usepackage[T1]{fontenc}

\usepackage{graphicx}	
\usepackage{bm}		
\usepackage{pdflscape}	
\usepackage[dvipsnames]{xcolor}
\usepackage{pifont}
\usepackage{tikz}

\newcommand{\simname}[1]{\texttt{#1}}

\title[Self-Sustaining Vortices]{Self-Sustaining Vortices in Protoplanetary Disks: Setting the Stage for \\
Planetary System Formation}

\author[Zs. Regály et al.]{
Zsolt Reg\'aly$^{1},$\thanks{E-mail: regaly@konkoly.hu}
Kundan Kadam,$^{1,2}$
and Cornelis P. Dullemond$^{3}$
\\
$^{1}$Konkoly Observatory, Research Centre for Astronomy and Earth Science, Konkoly-Thege Mikl\'os 15-17, 1121, Budapest, Hungary\\
$^{2}$Department of Physics and Astronomy, University of Western Ontario, London, Ontario N6A 3K7, Canada\\
$^{3}$Zentrum f\"ur Astronomie, Heidelberg University, Albert Ueberle Str. 2, 69120 Heidelberg, Germany
}

\pubyear{2021}

\begin{document}
\label{firstpage}
\pagerange{\pageref{firstpage}--\pageref{lastpage}}
\maketitle

\begin{abstract}
The core accretion scenario of planet formation assumes that planetesimals and planetary embryos are formed during the primordial, gaseous phases of the protoplanetary disk. However, how the dust particles overcome the traditional growth barriers is not well understood.
The recently proposed viscous ring-instability may explain the concentric rings observed in protoplanetary disks by assuming that the dust grains can reduce the gas conductivity, which can weaken the magneto-rotational instability. 
We present an analysis of this model with the help of GPU-based numerical hydrodynamic simulations of coupled gas and dust in the thin-disk limit.
During the evolution of the disk the dusty rings become Rossby unstable and break up into a cascade of small-scale vortices. 
The vortices form secularly stable dusty structures, which could be sites of planetesimal formation by the streaming instability  as well as direct gravitational collapse. 
The phenomenon of self-sustaining vortices is consistent with observational constraints of exoplanets and sets a favorable environment for planetary system formation.
\end{abstract}




\section{Introduction}

Over the past few decades, more than 3200 confirmed exoplanets have been detected \footnote{\href{http://exoplanets.org}{http://exoplanets.org}} and it seems that most low mass stars harbor planetary systems.
During the star formation process, protoplanetary disks (PPDs) are inevitably formed as a consequence of the conservation of angular momentum and these are considered to be the cradles of planets.
The widely accepted core-accretion paradigm considers the formation of such planets as a bottom up process, wherein the dust grains in the PPD grow into kilometer sized planetesimals \citep{Safronov1972}.
Collision of planetesimals leads to the formation of planetary embryos and subsequently terrestrial planets or giant planet cores \citep{KokuboIda1998}.
However, despite the immense progress in this field, how the small sized dust grains in the protostellar disk grow to become planetesimals is still an open question.
Several obstacles stand in the way which prevent the successful dust growth beyond approximately 1 meter in size.
Most notably, the drift barrier caused by aerodynamic drag on dust particles, results in their radial drift and eventual engulfment by the central star \citep{Weidenschilling1977}. The fragmentation barrier on the other hand tends to break up growing dust particles due to the large relative velocities \citep{Zsometal2010}.
The theoretical framework may be consistent, so long as some ``lucky few'' seed particles manage to escape these barriers to become planetesimals \citep{Windmark2012}. 

The role of vortices in the process of planet formation has been a subject of debate for a long time.
An anticyclonic vortex can be sustained in a Keplerian disk and it naturally forms a pressure maximum at its center.
As the dust particles are attracted towards the pressure maximum, vortices function as dust traps and can play an important role in planetesimal formation \citep{BargeSommeria1995,KlahrHenning1997}.
Simulations suggest that large-scale vortices can form in a PPD (e.g., at the outer boundary of the dead zone) and the resulting structures look remarkably similar to the asymmetries observed in sub-millimeter emission \citep{Regalyetal2012}. 
{The vortex formation can also occur due to the turbulence generated from vertical shear instability in PPDs \citep{Richard2016,MangerKlahr2018}}

Recently, \citet{DullemondPenzlin2018} performed a linear perturbation analysis and demonstrated that PPDs can be unstable to viscous ring-instability (VRI).
Small-sized dust grains tend to strongly reduce the conductivity of the gas by adsorbing electrons and ions \citep{Sanoetal2000,IlgnerNelson2006,Okuzumi2009}. 
This can inhibit the magneto-rotational instability (MRI) which is thought to be the primary source of turbulence in the disk \citep{BalbusHawley1998}. 
This can lead to a positive feedback loop as follows. 
An initial increase in the dust density reduces the MRI viscosity, which in turn can lead to accumulation of gas in its vicinity due to the mismatch in the mass transfer rate.
As the dust particles drift towards the resulting pressure maxima, they can amplify the initial perturbations into spontaneous formation of concentric rings \citep{DullemondPenzlin2018}. 
Thus, VRI may be a potential mechanism behind the grand design ring structures often observed in the sub-millimeter emissions of PPDs \citep{Andrewsetal2018,Dullemondetal2018}. 

In this paper, we present the results of the first coupled gas and dust simulations in the thin disk limit, that investigate the progression of the VRI with the help of 1D and 2D models.
In accordance with \cite{DullemondPenzlin2018}, the \citet{ShakuraSunyaev1973} $\alpha$ prescription for the disk viscosity was constructed such that the viscosity was suppressed with an enhancement in the dust-to-gas mass ratio.
Important disk physics such as gas and dust self-gravity, gas thermodynamics, and dust-feedback (with a fixed Stokes number, $\rm St$) were taken into account in the numerical simulations. 
We show that dusty rings formed via VRI become Rossby unstable and break up into a multitude of clumps, which develop into small-scale vortices.
The vortices formed are secularly stable and self-sustaining due to a positive feedback cycle similar to VRI.
As these vortices accumulate significant amount of dust, they set a favorable stage for formation of protoplanetary systems. 

The layout of our paper is as follows.
In Section~\ref{sec:hydro} we elaborate on the hydrodynamical model of the dusty protoplanetary disk and the initial conditions as well as the applied method for vortex analysis.
We discuss our major findings in Section~\ref{sec:results}, including the detailed analysis of three of the most interesting disk models. 
Finally, the implications of the results, model caveats and concluding remarks are presented in Section~\ref{sec:conclusions}.



\section{Hydrodynamical model}
\label{sec:hydro}
\subsection{Governing Equations}

We conducted the global 2D hydrodynamic simulations of PPDs using the the code {\small GFARGO2} \citep{RegalyVorobyov2017,Regaly2020}, which is our GPU supported version of the {\small FARGO} code \citep{Masset2000}.
In the gas-dust coupling, the dust was assumed to be a pressureless fluid. 
The dynamics of gas and dust components are described by the following equations:
\begin{equation}
  \frac{\partial \Sigma_\mathrm{g}}{\partial t}+\nabla \cdot (\Sigma_\mathrm{g} {\bm{v}})=0, 
  \label{eq:cont}
\end{equation}
\begin{equation}
  \frac{\partial \bm{v}}{\partial t}+(\bm{v} \cdot 
\nabla)\bm{v}=-\frac{1}{\Sigma_\mathrm{g}} \nabla P +\nabla \cdot \bm{\mathrm{T}} -  \nabla \Phi-\frac{1}{\Sigma_\mathrm{g}}\bm{f}_\mathrm{drag},         \label{eq:NS}
\end{equation} 
\begin{equation}
  \frac{\partial{e}}{\partial t}+\nabla \cdot (e \bm{v})=-P\nabla{\cdot 
  \bm{v}}+Q_\nu+Q_\pm,
  \label{eq:ENERG}\\
\end{equation}
\begin{equation}
  \frac{\partial \Sigma_\mathrm{d}}{\partial t}+\nabla \cdot 
    \Sigma_\mathrm{d} \bm{u}=-\nabla\cdot\bm{j}, 
    \label{eq:contd}
\end{equation}
\begin{equation}
   \frac{\partial \bm{u}}{\partial t}+(\bm{u} \cdot \nabla)\bm{u}=- \nabla\Phi+\frac{1}{\Sigma_\mathrm{d}}\bm{f}_\mathrm{drag}-(\bm{u}\cdot\nabla)\bm{j}, 
   \label{eq:NSd}
\end{equation} 
where $\Sigma_\mathrm{g}$, $\Sigma_\mathrm{d}$, and $\bm{v}$, $\bm{u}$ are the surface mass densities and velocities of gas and dust (being either dust particles or pebbles), respectively, and $e$ is the thermal energy density of the gas (per surface area). The gas pressure was given by
\begin{equation}
    P=(\gamma-1)e,
\end{equation}
where $\gamma=1.4$ is the adiabatic index of gas corresponding to molecular hydrogen. 

The gravitational potential ($\Phi$) of the system in Equations~(\ref{eq:NS}) and (\ref{eq:NSd}) is given by
\begin{equation}
\Phi(R,\phi)=-G\frac{M_*}{R}+\Phi_\mathrm{ind}(R,\phi)+\Phi_\mathrm{sg}(R,\phi),
\label{eq:phi_tot}
\end{equation} 
where the first term corresponds to the gravitational potential of the star in a given cell with radial distance $R$ and $G$ is the gravitational constant. Since the equations are solved in the cylindrical coordinate system centered on the star, Equation~(\ref{eq:phi_tot}) includes the so-called indirect potential, $\Phi_\mathrm{ind}(R,\phi)$ arising due to the displacement of the barycenter of the system caused by any disk non-axisymmetry (see its importance in, e.g., \cite{RegalyVorobyov2017}). The indirect potential is calculated as
\begin{equation}
\Phi_\mathrm{ind}(R,\phi) =  R\cdot G\int { dm(\bf{R}^\prime) \over R^{\prime 3} } \bf{R}^\prime,
\label{starAccel}
\end{equation} 
which in the cylindrical coordinate system can be given as
\begin{eqnarray}
\Phi_\mathrm{ind}(R_j,\phi_k) & = &  R_{j} \cos(\phi_k) \sum_{j',k'}G {m_{j',k'} \over R_j'^2 }\cos(\phi_{k'}) \nonumber \\
& & + \sin(\phi_k) \sum_{j',k'}G {m_{j',k'} \over R_j'^2 }\sin(\phi_{k'}),
\label{eq:phi_ind}
\end{eqnarray}
where $m_{j,k}$ and  $x_{j,k}$, $y_{j,k}$, are the mass and Cartesian coordinates of the grid cell $j,k$. To incorporate the effect of disk self-gravity, we calculate the gravitational potential of the disk, $\Phi_{\rm sg}$, by solving for the Poisson integral
\begin{eqnarray} 
  \Phi_\mathrm{sg}(R,\phi) & = & - G \int_{R_{\rm in}}^{R_{\rm out}} R^\prime dr^\prime 
                     \nonumber \\ 
      & &       \times \int_0^{2\pi} 
               \frac{\Sigma(R^\prime,\phi^\prime) d\phi^\prime} 
                    {\sqrt{{R^\prime}^2 + R^2 - 2 R R^\prime 
                       \cos(\phi^\prime - \phi) }}  \, ,
\end{eqnarray} 
where $R_{\rm in}$ and $R_{\rm out}$ are the radial position of the disk inner and outer boundaries. This integral is calculated using a FFT technique which applies the two-dimensional Fourier convolution theorem for polar coordinates logarithmically spaced in the radial direction \cite[see][Sect.\ 2.8]{BT87}.

In Equation~(\ref{eq:NS}) $\bm{T}$ is the viscous stress tensor of the gas, 
\begin{equation}
    \bm{\mathrm{T}} = \nu\left(\nabla \bm{v} + \nabla \bm{v}^T -\frac{2}{3}\nabla\cdot\bm{v}\bm{I}\right),
\end{equation}
whose components in polar coordinates are calculated according to \cite{Masset2002}. 
For the disk viscosity, $\nu$, we used the $\alpha$ prescription of \citet{ShakuraSunyaev1973}. 
In this case 
\begin{equation}
    \nu=\alpha \frac{c_\mathrm{s}^2}{\Omega_\mathrm{K}},
\end{equation}
where $\Omega_\mathrm{K}=(GM_*)^{1/2}R^{-3/2}$ is the Keplerian angular velocity, $M_*$ is the mass of the central star, and the sound speed. Assuming an ideal gas, $P=\Sigma c_\mathrm{s}^2/\gamma$, thus $c_\mathrm{s}$ can be given as 
\begin{equation}
    c_\mathrm{s}=\sqrt{\frac{\gamma(\gamma-1)e}{\Sigma_\mathrm{g}}}.
    \label{eq:cs}
\end{equation}
The $\alpha$ parameter characterizing the strength of the magneto-rotational instability 
was parametrized using the model described in \cite{DullemondPenzlin2018} as 
\begin{equation}
\alpha=\alpha_\mathrm{bg}\left(\frac{\Sigma_\mathrm{d}}{\Sigma^0_\mathrm{d}}\right)^{\phi_\mathrm{d}}\left(\frac{\Sigma_\mathrm{g}}{\Sigma^0_\mathrm{g}}\right)^{\phi_\mathrm{g}},
\label{eq:alpha}
\end{equation}
where $\Sigma^0_\mathrm{g}$ and $\Sigma^0_\mathrm{d}$ are the initial gas and dust densities. 
We limited the values of $\alpha$ between the background viscosity, $\alpha_\mathrm{bg}=10^{-2}$, and the MRI inactive viscosity, $\alpha_\mathrm{dead}=10^{-4}$. The latter is approximately the minimum effective viscosity arising due to vertical shear instability \citep{StollKley2016}. 
In the simulations described in this study, the values of $\phi_\mathrm{d}$ and $\phi_\mathrm{g}$ were set to -1 and 1, respectively, unless specified otherwise.
In this case, $\alpha$ was effectively inversely proportional to the dust to gas ratio, i.e., $\alpha \propto ({\Sigma_\mathrm{d}/\Sigma_\mathrm{g}})^{-1}$ or $\partial\mathrm{ln}\alpha/\partial\mathrm{ln}\Sigma=-1$. We also investigated a particular model where $\Phi_\mathrm{g}=\Phi_\mathrm{d}=-1$ was assumed.

The viscous heating ($Q_\nu$) was taken into to account as described in \cite{DAngeloetal2003}. Other sources of heating and cooling, such as stellar and background irradiation, were modeled implicitly via $Q_\pm$. 
We used $\beta$-cooling/heating prescription of \cite{LesandLin2015} to let the gas release/gain its internal energy assuming $\beta=1$, i.e., the e-folding timescale of disk cooling/heating to the initial temperature is one orbit at all distances.
The thermal evolution scheme can affect the gas dynamics \citep[e.g.][]{PierensLin2018}, however, this canonical prescription is sufficient for demonstrating formation and evolution of vortices.

The turbulent diffusion of dust is modeled by the 
gradient diffusion approximation (e.g., \cite{MorfillVoelk1984,Dubrulleetal1995,TakeuchiLin2002}), wherein the diffusive flux, $\bm{j}$, is given as 
\begin{equation}
    \bm{j}=-D\left(\Sigma_\mathrm{g}+\Sigma_\mathrm{d}\right)\nabla\frac{\Sigma_\mathrm{d}}{\Sigma_\mathrm{g}+\Sigma_\mathrm{d}}.
    \label{eq:turbdiff}
\end{equation}
According to \citep{YoudinLithwick2007}, the diffusion coefficient of dust is defined as
\begin{equation}
    D=\frac{\nu}{(1+\mathrm{St}^2)}.
    \label{eq:diffcoef}
\end{equation}
The Schmidt number, i.e., the ratio of momentum diffusivity (kinematic viscosity) and mass diffusivity, $\mathrm{S_c}=\nu/D$, is of the order of unity in our models.
According to \cite{Bentez-Llambayetal2019} the moment transported by the diffused material is taken into account in equation~(\ref{eq:NSd}).

Equations~(\ref{eq:NS}) and (\ref{eq:NSd}) and are solved by a two-step method. First, the source term, i.e. the right-hand sides are calculated then it is followed by the conventional advection calculation. For the source term, we use a fully implicit scheme (see details in \citealp{Stoyanovskayaetal2018}). The drag force exerted by the gas on the dust is calculated as
\begin{equation}
    \bm{f}_\mathrm{drag}=\frac{\bm{v}-\bm{u}}{\tau_\mathrm{s}},
    \label{eq:fdrag}
\end{equation}
where $\tau_\mathrm{s}=\mathrm{St}/\Omega$ is the stopping time. 
With this scheme, the effect of aerodynamic drag can be modeled for dust species that have stopping time that is much smaller than the time-step ($\tau_\mathrm{s}\ll\Delta t$). 
For pebbles that have large stopping time ($\tau_\mathrm{s}\gg\Delta t$), the method described is applicable as long as crossing orbits are not important for the dynamics. 
For simplicity, we assume that the dust has a fixed Stokes number  St=0.1 in models \simname{V1}-\simname{V3}.
In this case, at the inner edge of the disc ($R_\mathrm{min}=0.2$\,AU), the stopping time is $\tau_\mathrm{s}\simeq8.95\times10^{-3}$, while the time–step of the hydrodynamic simulation was measured to be $\Delta t=7.57\times10^{-4}$yr, thus $\tau_\mathrm{s}/\Delta t\simeq11.8$ there. 
Further out, e.g., at $R=1\,$\,AU, where the initial perturbation is applied, this ratio is about an order of magnitude larger.
To reveal the effects of dust size on the vortex self-replication, we run additional models assuming St=1 and 0.001.

The numerical resolution was $512 \times 1024$ cells in the radial (logarithmically distributed) and azimuthal (equidistant) direction, respectively. 
As a result, the resolution measured in the unit of local pressure scale-height is constant $0.125\,H$ throughout the disk.
At both the inner and outer boundary, the velocity components and gas surface density were damped to the initial value according to the method described in \cite{deValBorroetal2006}.

\subsection{Initial conditions}

In this study, we primarily focus on the results of six simulations as summarized in Table \ref{tbl:cases}.
We investigated the self-sustaining vortex formation in the terrestrial planet forming region of about 1 au  with model \simname{V1}. The outer disk at a distance of about 10~au was studied with model \simname{V2} and high metallicity environment was explored with model \simname{V3}.
The radial extent of the computational domain for the inner disk models (\simname{V1} and \simname{V3}) was between 0.2 au and 2.5 au, while for the outer disk (\simname{V2}), the extent was between 2 au and 25 au.
All simulations are evolved with the Stokes number of 0.1, except models \simname{V1(St=0.01)} and \simname{V1(St=1)}, which explore the effects of its variance.
Lastly, model \simname{V1($\phi_g=-1$)} investigates the magnetohydrodynamic (MHD) effects wherein the MRI activity can be increased in the gas depleted regions.
The initial distribution of gas was set to a power-law function of disk radius as
\begin{equation}
    \Sigma_\mathrm{g}^0=\Sigma_0 R^{-1/2},
\end{equation}
where the value at a distance of 1 au, $\Sigma_0=3900$ ${\rm g\, cm^2}$. 
In the limit of small disk mass (i.e., negligible disk self-gravity) such a disk is in quasi-steady-state, with a constant accretion rate throughout the disk.

In the analysis, two possible disk masses were investigated for each model by re-scaling $\Sigma_0$ to $2000$ and $4500$ ${\rm g\, cm^2}$, which covers the range for the canonical estimates for MMSN \citep{Adams2010}. 
The initial dust-to-gas mass ratio was set uniformly to 1\% for all models, except in the case of the high metallicity disk (\simname{V3}), where it was increased to 10\%.

We assume that the disk is initially geometrically flat, i.e., the disk pressure scale height is $H=h r$, where $h=0.05$ is the initial aspect ratio of the disk. Since $H=c_\mathrm{s}/\Omega$, 
\begin{equation}
c_\mathrm{s}^0=h r \Omega,
\end{equation}
and using Equation~(\ref{eq:cs}) the initial energy profile can be given as 
\begin{equation}
e^0 = \frac{(c_\mathrm{s}^0)^2 \Sigma_\mathrm{g}^0 }{\gamma( \gamma-1)}.
\end{equation}

In order to initiate VRI, an axisymmetric Gaussian perturbation was applied to the dust density at $1$ au, with an amplitude of 10 percent above the background density and a half-width at half maximum of 0.01 au.

\begin{table}
	\centering
	\caption{Summary of simulations showing self-sustaining vortex formation}
	\label{tbl:cases}
	\begin{tabular}{lcccl} 
		\hline
		{\bf Model name} & {\bf St} & {\bf \shortstack{$\phi_g$}} & {\bf \shortstack{$\phi_d$}} & {\bf \shortstack{Significance}} \\
		\hline
		\simname{V1} & 0.1 &  1 & -1 & inner disk (fiducial) \\
		\simname{V2} & 0.1 &  1 & -1 & outer disk \\
		\simname{V3} & 0.1 &  1 & -1 & high metallicity \\
		\simname{V1(St=1)} & 1 &  1 & -1 & grown dust \\
		\simname{V1(St=0.01)} & 0.01 &  1 & -1 & small-sized dust \\
        \simname{V1($\phi_g=-1$)}& 0.1 &  -1  & -1 & gas MHD effects \\
		\hline
	\end{tabular}
\end{table}

\subsection{Vortex analysis}
In order to quantify the collective behavior of the vortices formed within the disk, we used a standard algorithm 
to find the location and the area occupied by each vortex.
A vortex accumulates dust rapidly and thus creates an azimuthal asymmetry in the distribution of dust-to-gas mass ratio ($m_{\rm d}/m_{\rm g}$).
Thus, for detecting the location of all vortices at a given time, we identified local maxima in the field of $(m_{\rm d}/m_{\rm g}) / \langle m_{\rm d}/ m_{\rm g} \rangle$. 
Here, in order to avoid abnormally large averages in the presence of a vortex, the denominator was azimuthally averaged at the given radius over the lowest half of the values.
A region extending $10\times10$ grid cells in the $R-\phi$ plane was considered to have a local maximum if the difference between the maximum value and the minimum value within this region exceeds a certain threshold.
The optimal value of this threshold was set by eye for each simulation, 1.5 for model \simname{V2} and 0.4 otherwise.
If two vortices were in close proximity, i.e., if the distance between them was less than $0.2\times R$ and the radial separation $\Delta R < 0.05$ in the units of au, they were deemed to be multiple detection of a single vortex.
In such cases, the location of the maximum $\Sigma_{\rm d}$ was considered to be the center of the associated true vortex.
The area occupied by a vortex was considered to be an ellipse in cylindrical coordinate system, centered at its location.
The semi-major and semi-minor axes of such an ellipse were variable with radius, such that
$a_{\theta} = 0.24 (R/5)^{0.5}$ and $a_{R} = 0.03 (5/R)^{0.8} $, where $R$ is in the units of au.
The area thus chosen was conservative, in order to include all of the dust accumulated inside a typical vortex.
The vortices in close proximity with the inner as well as outer extent of the computational domain were rejected because of the boundary effects.
Note that the bulk properties of the vortices are not overly sensitive to the parameters chosen and we find this method to be sufficiently accurate to quantify the disk conditions.

The dust-to-gas volumetric density ratio is calculated by assuming vertical equilibrium
\begin{equation}
    \frac{\rho_\mathrm{d}}{\rho_\mathrm{g}}={\frac{1}{\sqrt{2\pi}}\Sigma_\mathrm{d}\frac{1}{H_\mathrm{d}}}/{\frac{1}{\sqrt{2\pi}}\Sigma_\mathrm{g}\frac{1}{H_\mathrm{g}}},
\end{equation}
where the gas pressure scale-height is
\begin{equation}
    H_{\mathrm{g}}=\frac{c_\mathrm{s}}{\Omega_\mathrm{K}}.
\end{equation}
To calculate the dust scale-height we assume size dependent vertical sedimentation for the dust, in which case,
\begin{equation}
    H_{\mathrm{d}}=H_\mathrm{g}\sqrt{\frac{\alpha}{\mathrm{St}+\alpha}}.
\end{equation}

To estimate the maximum size that solid particles could reach inside vortices during the dust growth process, we calculate the dust grain fragmentation according to \cite{Birnstieletal2012} as
\begin{equation}
a_{\rm frag} = \frac{2 \Sigma_{\rm g} v_{\rm frag}^2}{ 3 \pi \rho_{\rm s} \alpha c_{\rm s}^2}, 
\label{eq:afrag}
\end{equation}
with the typical assumptions of fragmentation velocity, $v_{\rm frag}=10$~m~s$^{-1}$, and the internal density of the dust aggregate, $\rho_{\rm s} = 1.6~\mathrm{g~cm}^{-3}$, with $\Sigma_{\rm g}$ calculated using the higher bound for MMSN.

\section{Results and Discussion}
\label{sec:results}
\subsection{Viscous ring-instability and Rossby wave instability}

\begin{figure*}
    \includegraphics[width=2\columnwidth]{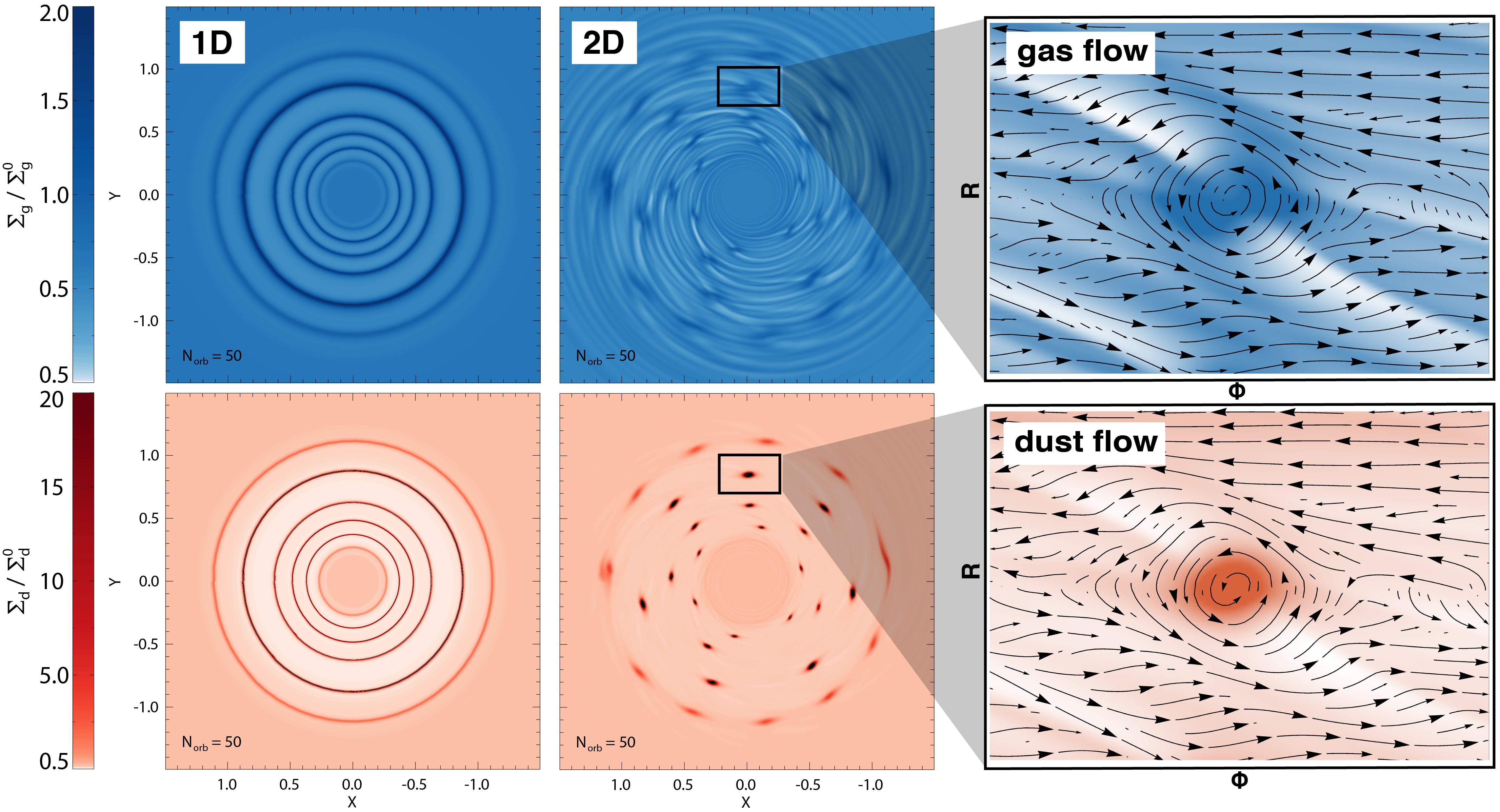}
    \caption{Excitation of viscous ring instability in 1D (first column) and Rossby instability in 2D for \simname{V2} model. Normalized gas and dust distributions at 50 orbits at 10 au are shown with blue and red colors. 
    Rings developed in 1D model become Rossby unstable and brake up and form small-scale vortices in 2D simulations which allow for non-axisymmetry. 
    The enlarged panels show gas and dust flow patterns in a frame of reference co-moving with the local Keplerian velocity, along with the respective surface density distributions.}
    \label{fig:1Dv2D}    
\end{figure*}

In this section we demonstrate the phenomenon of self-sustaining vortices qualitatively and elaborate on its general properties.
Fig.\,\ref{fig:1Dv2D} shows the contrast between 1D (neglecting the azimuthal components of hydrodynamic equations) and 2D simulations for model \simname{V2}. 
Consider the four panels on the left, showing the surface density distributions of dust and gas for the inner 15 au region of the disk.
Concentric rings formed in the 1D model due to VRI, while in the 2D case, the rings became Rossby unstable.
A representative vortex is magnified in the insets in the rightmost column.
The flow patterns show that the regions of enhanced gas and dust are indeed anticyclonic vortices.
The positive feedback loop responsible for VRI also makes the vortices self-sustaining and secularly stable.
As we shall see later, the vortices collect a significant amount of dust mass, while the midplane dust-to-gas density ratio grows beyond unity.
These conditions are ideal for rapid dust growth and the formation of planetesimals via the streaming instability \cite{Johansenetal2011}.
As we will see in Section \ref{sec:quanitative}, for particles with large Stokes number and in the outer disk, the central density in the vortices can exceed Roche density. Thus, the self-gravity can take over within such a vortex and initiate a direct gravitational collapse.
.

\begin{figure*}
    \includegraphics[width=2.1\columnwidth]{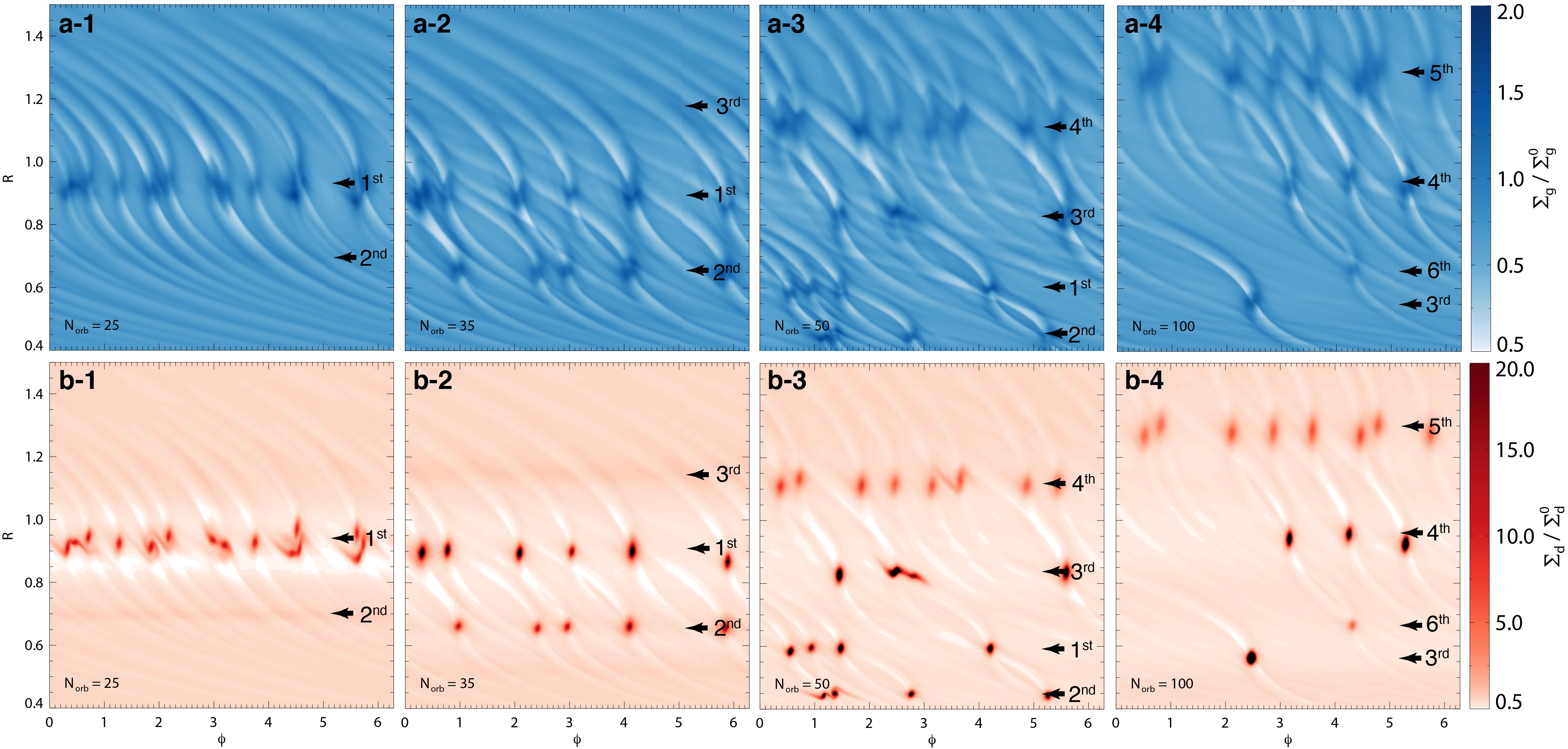}
    \caption{Progression of self-sustaining vortex cascade in \simname{V1} model. 
    The fields are sequential snapshots of the surface density distributions of gas (top) and dust (bottom), normalized with respect to the initial values and plotted in polar coordinates. 
    The numbers on the right indicate the generation of the vortices, with the arrow pointing at their approximate radial position. 
    The formation and orbital decay of successive generations as well as instances of vortex merging can be noticed.}
    \label{fig:gas-dust-RWI}    
\end{figure*}

In all simulation setups, we find that the protoplanetary disk exhibited VRI near the initial dust density perturbation, followed by development of RWI in the non-linear regime and vortex formation\footnote{The onset of RWI (\simname{modV1-early.m4v}) and dynamics of matured vortices (\simname{modV1=late.m4v}) recorded in short movies available in the on-line version. Movies show the distribution of the dust component in a frame that co-rotates with the disk at $R = 1$, where the initial perturbation is applied.}.
We demonstrate the progression of this behavior qualitatively in Fig.\,\ref{fig:gas-dust-RWI} with the help of simulation \simname{V1}.
The figure shows the evolution of gas and dust density distributions in polar coordinates ($R$-$\phi$ plane). 
Due to the positive feedback loop described earlier, the VRI was set in motion in the vicinity of the initial dust perturbation applied at 1 au. 
The VRI manifested as an axisymmentric ring in both dust and gas, increasing in magnitude over time, while also migrating inward due to the viscous evolution of the disk.
In model \simname{V1}, VRI developed over the first 20 orbits ($N_{\rm orb}$).
When the radial pressure gradient in this ring became sufficiently large, RWI was triggered. 
Panels a-1 and b-1 in Fig.~\ref{fig:gas-dust-RWI} show the instance when the VRI became nonlinear and RWI clearly appeared in the simulation.
A ring initially fragmented into dusty clumps with a typically large azimuthal mode number ($m \simeq 20$).
The dust accumulated at a faster rate than the gas, due to its strong affinity towards pressure maxima (see panel b-1).
Panels a-2 and b-2 of Fig.\,\ref{fig:gas-dust-RWI} show the progression of RWI into several well-defined, anticyclonic vortices.
The measured aspect ratio (the radial extent divided by the azimuthal extent) was found to be about 0.45 for a fully formed, isolated vortex. The typical size of a vortex was 0.13 au at 1au (about twice $H_{\rm g}$) and it increased with the radius.
Since the temperature change inside a vortex was negligible due to its relatively small size, the local pressure scale height did not change appreciably.

We emphasize that the vortices were self-sustained in such a way that the positive feedback loop between the dust accumulation, reduced viscosity, and the gas accumulation makes them secularly stable in a background of high viscosity.
Note that in previous studies without this effect, the canonical values viscosity ($\alpha\approx10^{-2}$) as well as the backreaction of dust particles tended to smear out and destroy the vortices formed in a PPD \citep{Fuetal2014}.
The primary interaction between vortices was merger and it was reflected in a reduction in their number over time.
The vortices were typically stable over thousands of orbits until they encountered the inner boundary. 
The gas distribution showed large scale spiral waves which originated at the location of each vortex.
Because of the smaller mass accumulation, the spirals excited by these vortices were much weaker than those induced by large-scale vortices in pure gas simulations (see, e.g., \citealp{RegalyVorobyov2017} and references therein).
Rossby vortices have been shown to drive angular momentum transport in the disk \citep{Paardekooperetal2010}.
The resulting viscous torques can cause their orbital decay or inward migration, however, this phenomenon warrants further investigations.

The large scale spiral density waves associated with these vortices have a very interesting consequence.
The waves can interfere constructively at a different radius, either inside or outside, giving rise to perturbations in gas surface density.
With the positive feedback loop mentioned above,
these gas enhancements can collect dust and result in a vortex cascade, or the formation of a new ``generation'' of vortices. 
Thus, we term the original set of vortices formed at the location of the ring-instability as the first generation.
Panels a-2 and b-2 of Fig.~\ref{fig:gas-dust-RWI} depict the formation of a second and third generations of vortices, clustered at the radius of about 0.6 and 1.2 au.
The last two columns of Fig.~\ref{fig:gas-dust-RWI} show maturation of the second and third generation vortices after mergers and inward migration, as well as formation of next generations of vortices.
Note that it is also possible that VRI can form multiple rings first, which subsequently become Rossby unstable.

\subsection{Vortex evolution in terrestrial planet forming region}
\label{sec:quanitative}

\begin{figure*}
\hspace{0.8cm} \simname{V1}  \hspace{5cm} \simname{V2}  \hspace{5.2cm} \simname{V3}  \\
    \includegraphics[width=0.65\columnwidth]{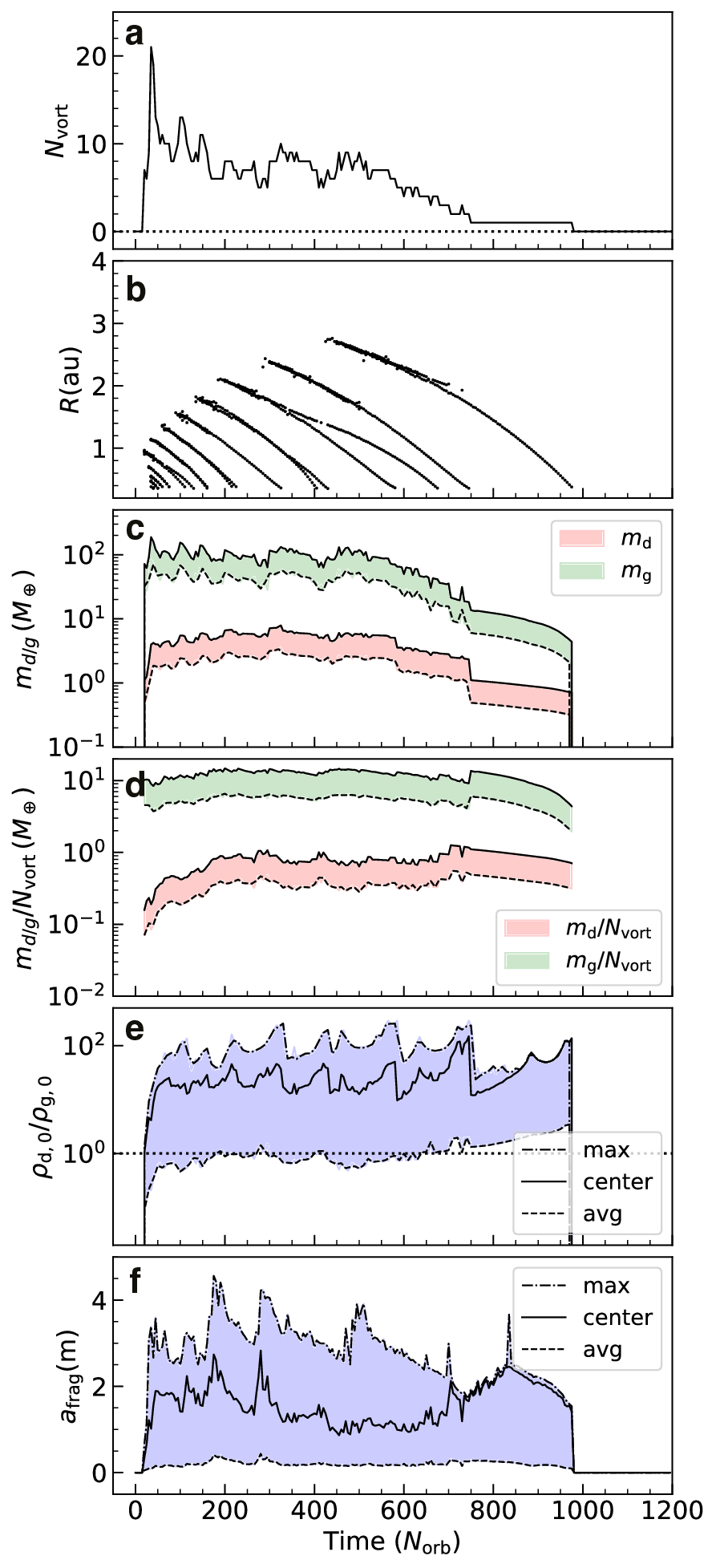}
    \includegraphics[width=0.65\columnwidth]{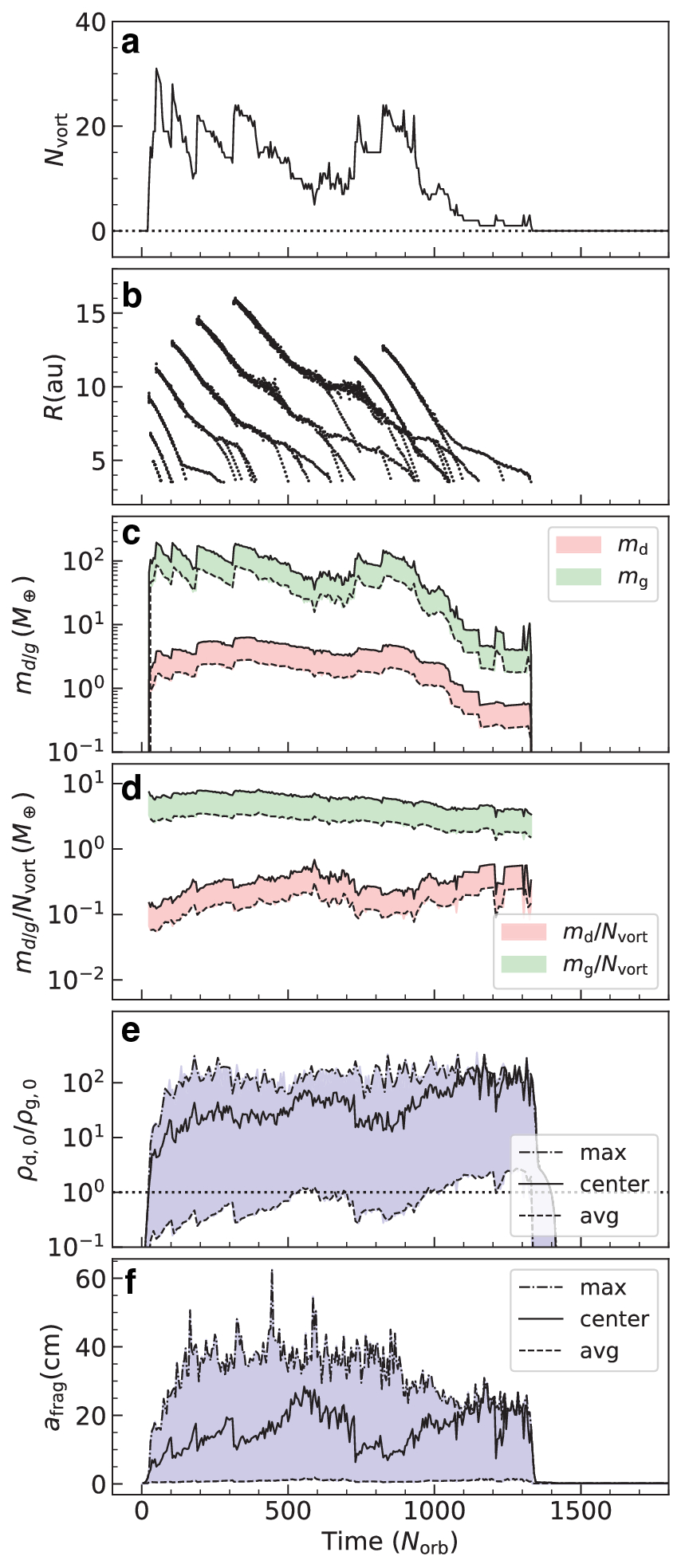}
    \includegraphics[width=0.65\columnwidth]{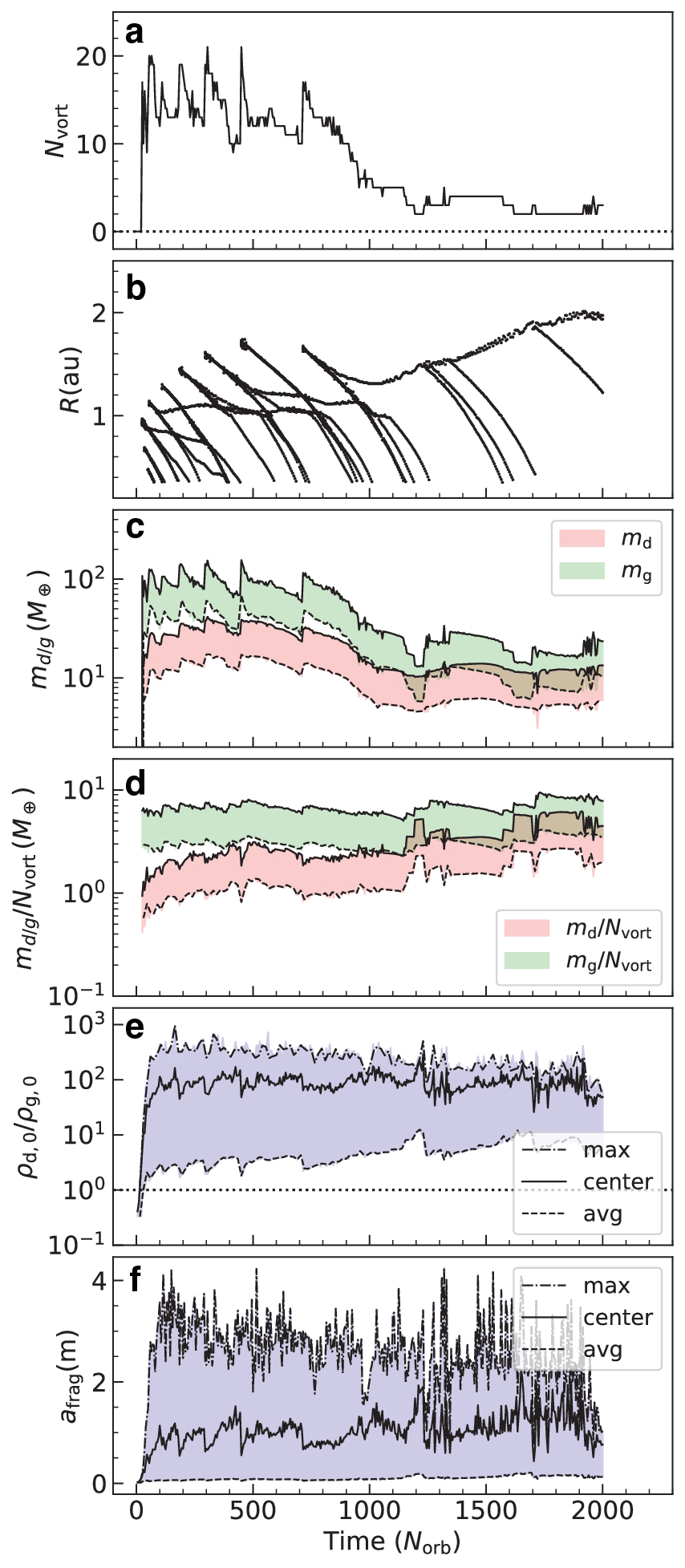}
    \caption{ Temporal evolution of the aggregate properties of the vortices formed in models \simname{V1} (terrestrial planet forming region), \simname{V2} (outer disk) and \simname{V3} (higher metallicity), conducted with St=0.1. 
    The rows show the following time-dependent physical parameters: (s) the number of vortices identified in the disk; (b) spacetime diagram of the radial distance evolution of the vortices; (c) the minimum and the maximum bounds of the total dust and gas mass inside vortices; (d) the average amount of dust and gas mass inside an individual vortex;
    (e) dust-to-gas volumetric density ratio at the disk midplane; (f) dust grain fragmentation radius. 
    The shaded regions in panels~c and d correspond to the maximum/minimum values calculated using the upper/lower bounds of gas density for the MMSN model. The dotted lines in panels~e and f show the excitation criterion for triggering streaming instability and the meter size, respectively.}
    \label{fig:evol_St01-C1}    
\end{figure*}

Consider the first column in Fig.\,\ref{fig:evol_St01-C1}, depicting the temporal evolution of the disk with respect to the vortex properties for \simname{V1} model. 
The simulation focuses on the region in the vicinity of 1~au, which is the terrestrial planet forming region of a PPD. 
Panel~a shows the total number of vortices identified on the disk ($N_{\rm vort}$) as a function of time, while panel b shows the spacetime diagram of the radial distance evolution of each vortex.
The multiple generations of vortices can be clearly seen in panel~b, 
each corresponding to an abrupt increase in $N_{\rm vort}$.
The reduction in $N_{\rm vort}$ over time is caused by vortex coagulations or mergers.
As the disk evolved, the dust particles drifted inward and left the computational domain through the inner boundary.
In addition, the dust was also collected by the vortices, resulting in a decline of the background dust content of the disk.
Thus, the production of vortices stopped at about 500 orbits.
The average rate of migration of the vortices ($\sim 5 \times 10^{-3}$~au~yr$^{-1}$) was about an order of magnitude faster than the calculated dust drift speed at 1 au.

Panel~c for the model \simname{V1} in Fig\,\ref{fig:evol_St01-C1} shows the total mass of dust as well as gas accumulated in all the vortices, while panel~d shows the same quantities averaged for a single vortex.
The colored area between solid and dashed lines encompasses the maximum and minimum estimates of mass, respectively, based on the bounds for MMSN.
Despite of the migration of the vortices, $m_{\rm g}$ per vortex remained remarkably constant for \simname{V1} model.
The average dust mass in a vortex increased marginally over time in the vicinity of about 0.1~$M_\oplus$. 
Note that the evolution described here reflects a combination of events for a number of vortices. 
As soon as a vortex is formed with RWI, it starts collecting dust, while dust mass is lost during its inward migration via the process of diffusion.
During the merger of two vortices, the dust mass is approximately conserved.
However, the resulting vortex looses dust at an increased rate, thus maintaining a balance between dust drift towards the pressure maximum in the vortex and its diffusion because of the increased dust surface density.
Note that although the ratio of dust-to-gas surface density may increase by a factor of two or up to an order of magnitude during a merger, it always remained below unity.

Panel~e shows the midplane volumetric dust-to-gas density ratio ($\rho_{\rm d}/\rho_{\rm g}$) assuming vertical equilibrium in gas and solid settling.
The ``max'' and ``center'' values are the maximum and average values of this quantity at the vortex location, respectively.
The ``avg'' values are averaged over the vortex area and are thus the lower bounds.
Here the blue color encompasses the area between the estimated maximum and minimum values. 
As seen in panel~e, the midplane dust density can be over 10 times larger than the gas inside a vortex.
The corresponding dust-to-gas mass ratio often exceeded unity at the center of the vortices, indicating significant dust accumulation.
Thus, although the vortices migrated at a faster rate, the local conditions could trigger streaming instability near the vortex centers and assist planetesimal formation \citep{YoudinGoodman2005}.

The last panel~f shows the dust grain fragmentation radius ($a_{\rm frag}$) calculated with the typical assumptions for PPD \citep{Birnstieletal2012} and using the higher mass bound for MMSN (Equation \ref{eq:afrag}).
The dust pebbles can achieve an approximate maximum size of $a_{\rm frag}$ before they are destroyed through fragmentation resulting from high velocity collisions.
With $a_{\rm frag}$ often exceeding 1~m, the vortices offer a more favorable environment for planetesimal formation, as compared to the traditional disk models \citep{HengKenyon2010}.

\subsection{Vortex evolution in the outer disk}

A PPD can show significantly different behavior in the outer regions, where the disk self-gravity can dominate and affect its dynamical evolution.
A sufficiently massive disk is prone to gravitational instability and fragmentation into gaseous clumps, due to the low values of Toomre's Q-parameter at large radii.
In order to study the vortex cascade in the outer disk region, we conducted simulation \simname{V2} in the vicinity of a radial distance of approximately 10~au from the central star. 
In the case of model \simname{V2}, the Q-parameter dropped to $\sim2$  near the vortices, implying that the self-gravity was significant in these regions. 
However, note that the Q-parameter remained above unity throughout this, as well as all simulations described in this study, indicating that the disks always remained gravitationally stable.
We found that the disk self-gravity renders vortices more regularly spaced than in low-mass disk models (see Fig.~\ref{fig:1Dv2D}).
The aspect ratio of an individual vortex remained similar to model \simname{V1} (about 0.45), however, because of their radial location, the vortices were about ten times larger at 10 au as compared to 1 au.

\begin{figure*}
    \includegraphics[width=2\columnwidth]{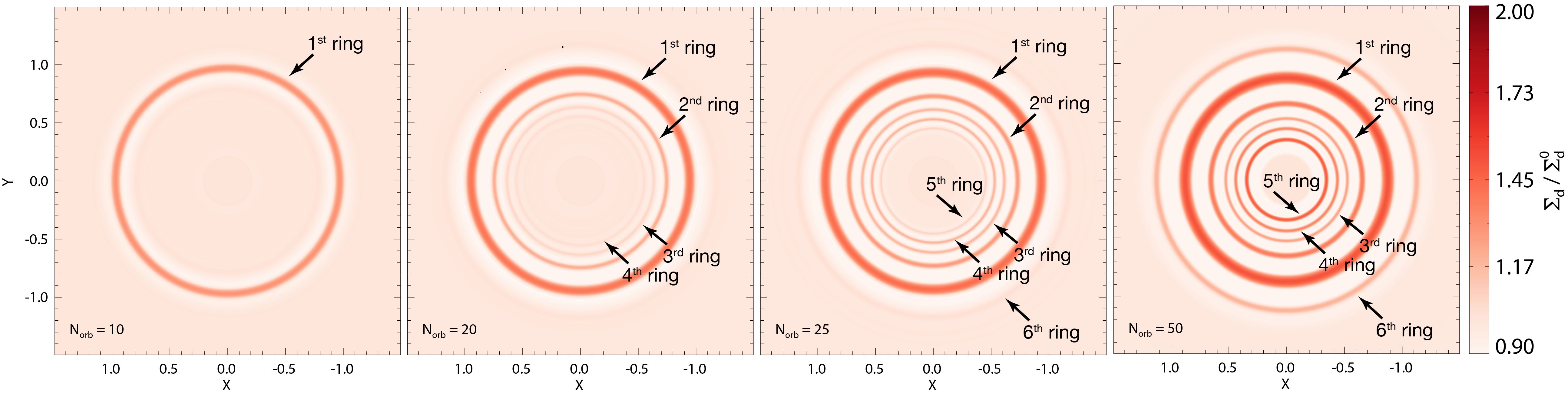}
    \caption{Formation and development of dusty ring structures in the model with St=0.1 solid species (normalized with the initial surface mass density) via VRI without the excitation of RWI. 
    The excitation of RWI was suppressed in this model by setting $\alpha_\mathrm{min} / \alpha_\mathrm{bg} = 0.9$. 
    It is appreciable that the rings of solid material are subject to inward drift as they evolve.}
    \label{fig:evol_St01-C1b_r}    
\end{figure*}

Consider the second column in Fig.\,\ref{fig:evol_St01-C1}, which shows the temporal evolution of model \simname{V2}.
As seen in panel~a, the number of vortices formed in each generation was about twice as many as compared to model \simname{V1}.
The inward migration rate of the vortices in the outer disk was about an order of magnitude slower ($\sim 2 \times 10^{-5}$~au~yr$^{-1}$) as compared to the dust drift speed at the radius of 10 au.
Thus, these vortices can offer ideal sites for overcoming the dust drift barrier.
The amount of gas as well as dust confined by vortices formed at the outer disk was similar to that found in terrestrial planet forming region (see panels~b and c).
As compared to the inner disk, the dust-to-gas volumetric density ratio (panel~e) was marginally larger in the outer than in the terrestrial planet forming region of PPD.
The maximum fragmentation size of the dust grains in the outer disk reached about half a meter.
We hypothesize that both the increased number of vortices as well as the enhanced concentration of the dust were assisted by the disk self-gravity.

Note that the development of VRI in the disk does not always lead to vortex formation.
At low densities expected in the outer disk, the ionization degree is much less sensitive to the dust properties \citep{Dzyurkevich13}.
When this reduced sensitivity to the dust content is modeled with a larger value of the minimum viscosity, i.e. $\alpha_\mathrm{min} / \alpha_\mathrm{bg} \gtrapprox 0.7$, the vortex formation is suppressed and only VRI is excited without the secondary RWI.
Fig. \ref{fig:evol_St01-C1b_r} shows a particular simulation in which dust with St=0.1 was modeled in the inner disk assuming a similar parametric-$\alpha$ model, but with $\alpha_\mathrm{min}  = 0.9 \alpha_\mathrm{bg}$.
A series of dusty rings were formed and continued to evolve over a large number of orbits without the excitation of RWI.
 The instability grows the quickest for the shortest wavelength, which is the disk pressure scale height.
In general, the width of the substructure was consistent with this length-scale and the value of $\Delta r/r$ was between 0.05 and 0.5, which is comparable to the typical values observed in PPDs \citep{Huang2018}.
However, the gaps in the simulations were consistently wider than the rings.
For a particular phase shown on the last panel of Fig.~\ref{fig:evol_St01-C1b_r}, the widths were found to be $\Delta r/r \simeq 0.15$ and $0.08$ at 1 au for the gaps and the rings, respectively.
Similar to the small-scale vortices described earlier in models \simname{V1}-\simname{V3}, the rings were also subject to an inward migration.
This demonstrates that the observations of grand-design, concentric ring-like structures in sub-millimeter wavelength emission (e.g., \cite{Dullemondetal2018}) can be accounted for within this framework.

\subsection{Effect of stellar metallicity}

One of the major observations related to exoplanets statistics is that the frequency of giant planets is strongly correlated with the metallicity of the host star, presumably because of an enhanced rate of planetesimal formation in the disk \citep{Gonzalez97,FischerValenti05, Johansen09}.
We investigated the effect of stellar metallicity on the vortex formation with model \simname{V3}, assuming 10 times greater initial dust mass distribution.
Note that the limits of the effective $\alpha$ parameter were similar to the previous simulations, since this enhanced initial dust density ($\Sigma_{\rm d}^{\rm 0}$) was used for its calculation (Equation \ref{eq:alpha}).
Consider the third column in Fig.~\ref{fig:evol_St01-C1} showing the temporal evolution of this high metallicity PPD.
The dust back-reaction is known to shorten the vortex life-time \citep{Fuetal2014}, however,
the stronger drag forces associated with higher dust content did not inhibit self-sustaining vortex formation (panel~a) or shorten vortex lifetime (panel~b).
Both the total mass contained in the vortices as well as the average mass within a single vortex were about 10 times larger as compared to the solar metallicity counterpart (panels~c and d).
Similarly, the corresponding midplane dust-to-gas density ratio was an order of magnitude greater on average.
Due to this larger density ratio, the viscosity in this region was also proportionally lower.
The lower viscosity can explain the inefficiency of dust back-reaction in shortening the vortex lifetime.
The fragmentation size of the dust particles was found to be about twice as large compared to the standard \simname{V1} model (panel~e).
Thus, the dynamics of the self-sustaining vortices was strongly affected by the stellar metallicity.
The vortices also became more compact with an increase in metallicity and a typical vortex was about 0.1 au across at 1au.

An interesting phenomenon was observed in model \simname{V3}, wherein a mature vortex splits in two vortices during its evolution. 
The new vortex migrates inward while the progenitor migrates in the outward direction.
This can be noticed in panel~b of Fig.~\ref{fig:evol_St01-C1}, with a  forked distance evolution beyond about 1000 orbits.
The total dust mass within a vortex is approximately conserved during such a split and the individual daughter vortices then start gaining dust mass.
A preliminary analysis indicates that such vortex splitting occurs due to the PPD's self-gravity, for reasons analogous to the phenomenon of vortex stretching \cite{RegalyVorobyov2017}.

\subsection{Vortex formation in the early and late stage PPDs}

\begin{figure*}
\begin{flushleft}
\hspace{3cm} \simname{V1(St=1)}  \hspace{3.5cm} \simname{V1(St=0.01)}  \hspace{3.8cm} \simname{V1($\phi_\mathrm{g}=-1$)}
\end{flushleft}
    \includegraphics[width=0.65\columnwidth]{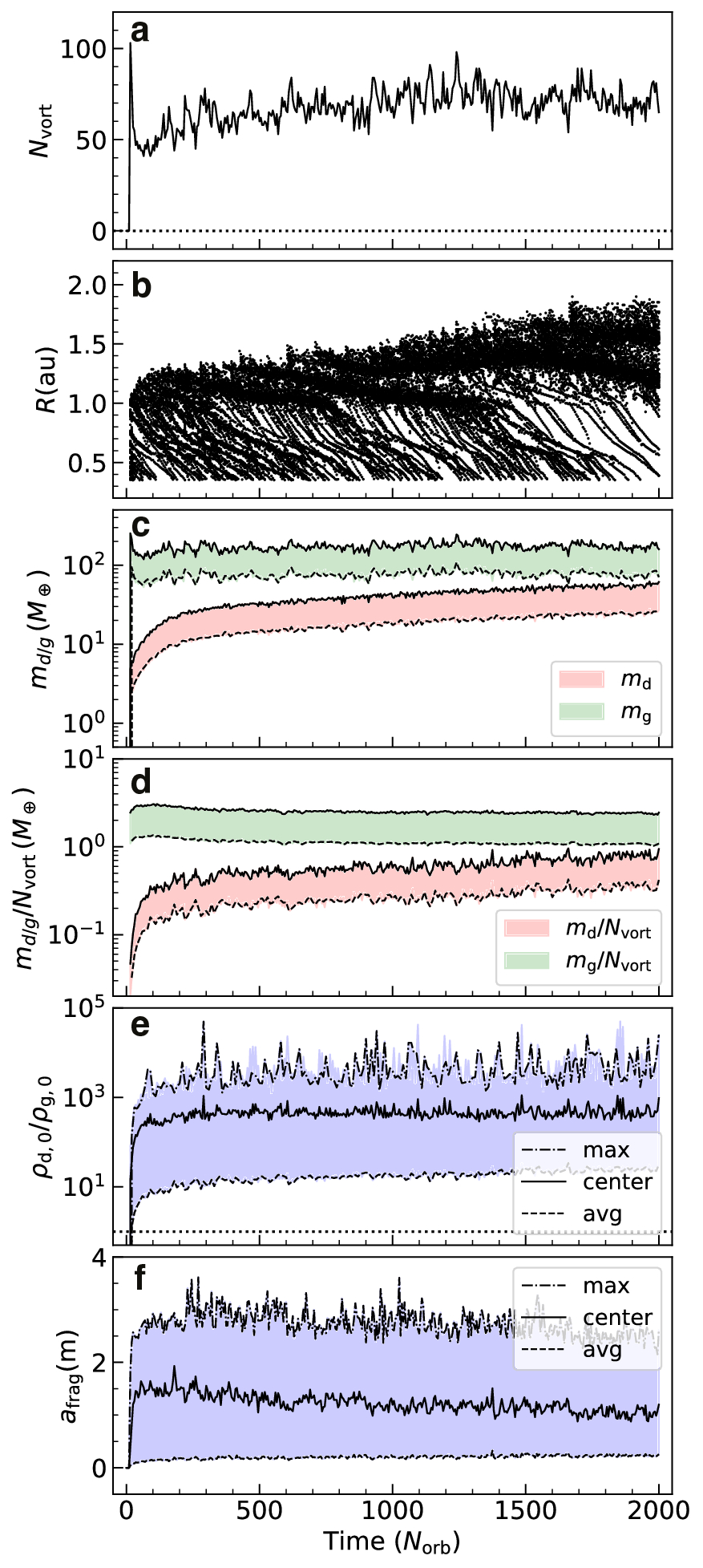}
    \includegraphics[width=0.655\columnwidth]{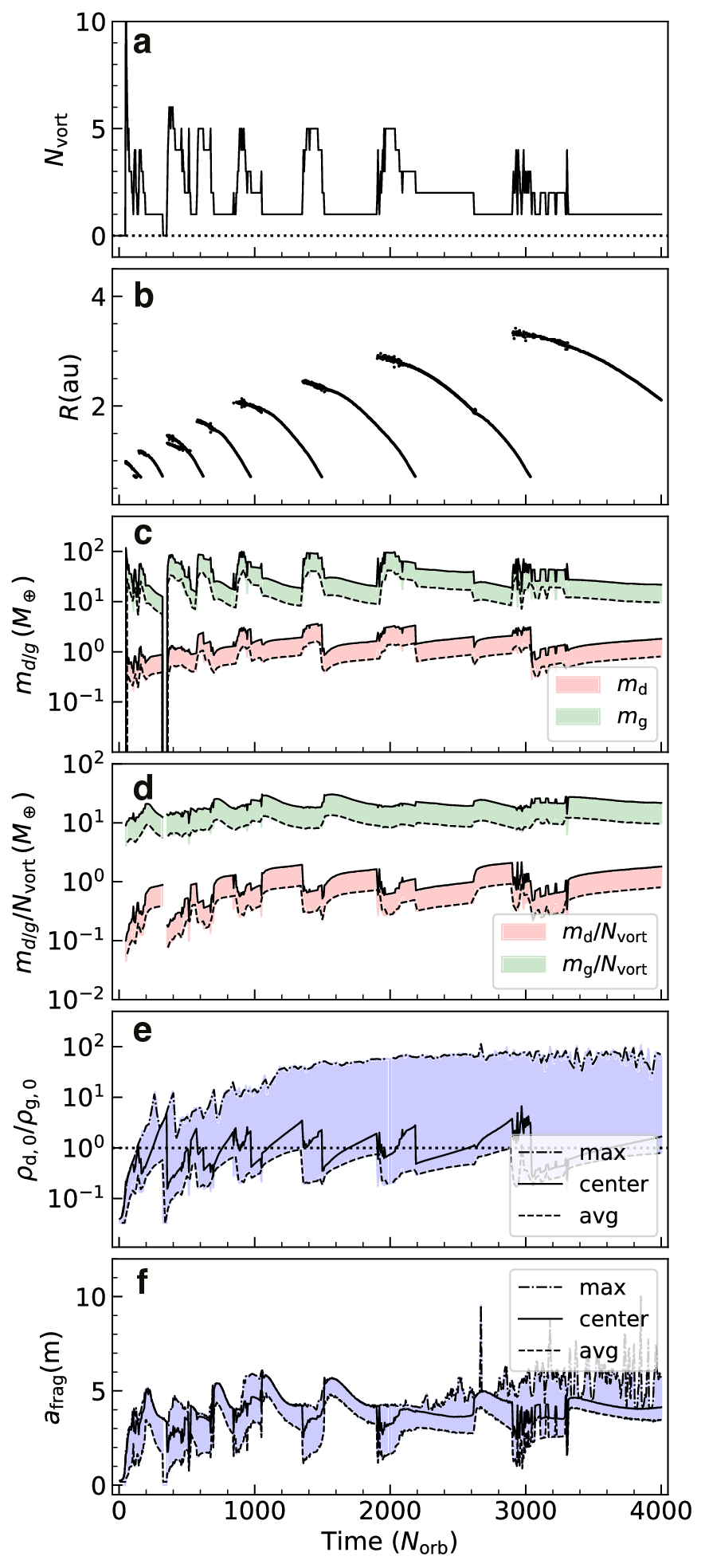}
    \includegraphics[width=0.66\columnwidth]{St01-c1b_final.jpg}
    \caption{Temporal evolution of the aggregate properties of the vortices formed in modified models \simname{V1} (terrestrial planet forming region). Simulations conducted with St=1 and St=0.01 for the first two columns, respectively. Right column shows the model where $\phi_\mathrm{g}=-1$ and St=0.1 are assumed.
    Individual panels a)-f) show the same time-dependent parameters as in Fig.\,\ref{fig:evol_St01-C1}.}
    \label{fig:evol_St1-001-C1-c1}    
\end{figure*}

As a PPD evolves, the dusty component of disk is subject to growth process, forming progressively larger sized particles.
For investigating self-sustaining vortex formation in disks at different evolutionary stages, we conducted additional simulations (analogous to \simname{V1} model) assuming larger and smaller dust component with Stokes numbers of 1 and 0.01, respectively.
In order to excite RWI, a strong pressure gradient is required, which is reinforced by the dust enhancement in our model (see Equation \ref{eq:alpha}). 
On the other hand, the turbulent diffusion of the dust component tends to inhibits its accumulation.
The Stokes number of the dust particles directly affects both the diffusion coefficient as well as the dust back-reaction (Equations \ref{eq:diffcoef} and \ref{eq:fdrag}, respectively), and thus has a significant impact on vortex formation.

For the model \simname{V1(St=1)}, the turbulent diffusion of solids was practically negligible {see the first column in Fig.\,\ref{fig:evol_St1-001-C1-c1}).
This resulted in an enhanced vortex cascade and the number of vortices form was an order of magnitude larger ($\sim 150$) than in \simname{V1} model, while the duration of the vortex phase was about twice as long.
This indicates that the efficiency of the vortex formation process is enhanced for large-sized dust particles.

In the simulations presented so far, the central density of the vortices was well below Roche density. For model \simname{V1}, the central density was less than $1\%$ of the Roche density. The maximum ratio of central density to Roche density increased in the outer disk ($10\%$ for model \simname{V2}) and at an increased metallicity ($5\%$ for model \simname{V2}). 
However, at an increased Stokes number in model \simname{V1(St=1)}, the central density of a vortex can reach about $40\%$ of the Roche density at that radius. 
Such a large value resulted in significant effects due to self-gravity and the “vortices” tended to loose a well-defined shape and were deformed into dusty clumps.
In an outer disk simulation with St=1 (not presented, since the evolution was similar to \simname{V1(St=1)} with trends that are already discussed), the central density can occasionally exceed Roche density. 
This is significant as it implies that the vortices can possibly show direct gravitational collapse in the outer disk regions.
This in unlike the large-scale vortices that tend to stretch in massive discs, which prevents them from gravitational collapse \citep{RegalyVorobyov2017}.

In the case of St=0.01 and 0.001 models, where the dust particles are well coupled with the gas, self-sustaining vortices also developed. 
For model \simname{V1(St=0.01)} shown in the second column of Fig. \ref{fig:evol_St1-001-C1-c1}, the vortex migration was slower and the vortex generation was extended to about four times longer than St=0.1 case.
However, the vortex formation either required a lower background viscosity, e.g., $\alpha_\mathrm{bg}=10^{-3}$ or a stronger dust sensitivity of viscosity, (i.e., $\phi_\mathrm{d}=-2$, see Equation~\ref{eq:alpha}).
The former can be explained by the fact that the strength of dust diffusion is proportional to the background viscosity (Equations~\ref{eq:turbdiff} and \ref{eq:diffcoef}).
Extended regions of low effective viscosity can be found in the dead zone of a PPD where this condition may be satisfied.
The small-sized dust grains can exhibit stronger recombination efficiency provided by the increased surface area per unit mass.
Thus, a steeper dependence of viscosity on the dust content, i.e., a larger magnitude of $\phi_\mathrm{d}$, is indeed possible at small Stokes number.
The reluctance of lower Stokes number models to form vortices also suggests that the dust diffusion works against vortex formation.

\subsection{Emulating non ideal MHD effects}
\label{sec:MHDeff}

Local shearing-box simulations of a PPD in the non-ideal MHD regime show that 
MRI turbulence can generate long-lived, large-scale, axisymmetric banded density variations which are enhanced by an external magnetic flux \citep{BaiStone2014}.
In these ``zonal flows", the vertical magnetic flux is strongly enhanced or diminished to nearly zero in the low or high density regions, respectively \citep{Bai2015}. 
Enhanced magnetization in the zonal flows can cause local changes in the MRI such that the effective viscosity increases in the low density gaps and is suppressed in the denser regions.
Thus multiple gaseous rings can form in the disk independent of its dust content.
The dust collecting efficiency of such zonal flows, however, has shown to be limited in the outer disk and they are unlikely to lead to significant particle trapping \citep{SimonArmitage2014}. 

To model layered accretion, \citet{HasegawaTakeuchi2015} derived an empirical formula for Shakura \& Sunyaev $\alpha$-parameter being proportional to $\Sigma_g^0/\Sigma_g$, which is equivalent to $\phi_g=-1$ in our model (see Equation~(\ref{eq:alpha})). 
We can mimic the zonal flows with this prescription, as the MRI activity is inversely proportional to the gas accumulation.
Note that the models discussed so far in this paper assume $\phi_g=1$, for which case dust drift is strictly required for the VRI to operate \citep{DullemondPenzlin2018}. 
To investigate the consequences of MHD effects on ring-instability and subsequent Rossby wave instability, we conducted an additional simulation \simname{V1($\phi_g=-1$)} which is otherwise identical to model \simname{V1}.

In the case of model \simname{V1($\phi_g=-1$)}, we found that the VRI is excited and these rings become Rossby unstable as in previous models. Moreover, the small scale vortices are also subject to a coagulation process. Analysis of the results are shown in the third column of Fig.\,\ref{fig:evol_St1-001-C1-c1}. The production of vortices is somewhat extended to a larger distance compared to our reference model \simname{V1}. The total and the average dust and gas mass collected by the vortices (thus the enhancement in dust-to-gas mass ratio) are slightly enhanced in model \simname{V1($\phi_g=-1$)}, although the dust grain fragmentation radius shows a similar evolution to model \simname{V1}.
The relative insensitivity of self-sustaining vortex formation on $\phi_g$ implies that this phenomenon is primarily governed by the dust dynamics.

\section{Conclusions}

\label{sec:conclusions}
We studied a physically motivated model of PPD evolution wherein the MRI viscosity in the disk was suppressed with the local enhancement in the dust-to-gas ratio.
The numerical experiments were carried out with the help of 2D global hydrodynamic simulations, which included the dust dynamics as well as gas-dust coupling.
We presented the detailed results for three scenarios- terrestrial planet forming region, the self-gravitating outer disk and a high metallicity PPD.
We show that when the viscosity depends on the dust-to-gas mass ratio,
the concentric rings formed due to viscous ring-instability \citep{DullemondPenzlin2018} become Rossby unstable and 
as a result, the dusty ring breaks up into a multitude of small-scale, self-sustaining vortices.
The vortices are secularly stable due to the positive feedback loop responsible for VRI and are subject to complex interactions such as mergers and orbital decay. 
An individual vortex could collect a significant amount of dust (0.1-10$M_\oplus$). 
A number of traditional barriers to the dust growth and planetesimal formation can be overcome inside the vortices due to the suitable local conditions as well as their stability over large number of orbits.
Moreover, the vortices can possibly collapse due to their self-gravity in the outer disk region.
Thus, the phenomenon of self-sustaining vortices sets a favorable stage for the formation of planetary systems with the core accretion scenario.
Recent measurements show that protoplanetary disks around class II pre-main sequence stars do not contain enough dust mass needed to form the observed populations of exoplanet systems \citep{Tychoniec2020}. 
This puts severe constraints on the timescale and efficiency on the planet formation process and the phenomenon of self-sustaining vortices may be a crucial piece of the puzzle.  

We emphasize that this is the first study to show that a large number of stable vortices can continually form in a PPD, as long as the dust is replenished from the outer regions.
This phenomenon is consistent with several observational constraints such as enhanced planet formation in high metallicity PPDs \citep{FischerValenti05} and high efficiency of the planet formation process \citep{Tychoniec2020}.
The self-sustaining vortices not only offer suitable conditions for planetesimal formation but also point towards the tantalizing idea of rapid, in situ formation of planetary cores.

As a final remark, in this study we used fixed Stokes numbers for the dust and therefore the processes of dust growth or fragmentation were neglected.
Furthermore, the surface area available for ion and free electron recombination is smaller for grown dust as compared to small sized dust, and thus, the value of $\phi_\mathrm{d}$ in equation\,(\ref{eq:alpha}) can be size-dependent. 
Another theoretical challenge is that our study did not include an ab initio treatment of the ionization physics and should be followed up with comprehensive models containing relevant processes.
Understanding the details of the vortex behavior such as their migration, merger and splitting properties also warrants further investigation.

\section*{Acknowledgements}

The project was supported by the Hungarian OTKA Grant No. 119993. 
K.K. acknowledges funding from the European Research Council (ERC) under the European Union's Horizon 2020 research and innovation program, grant No. 716155 (SACCRED). We thank the referee for her/his helpful comments that significantly improved the quality of the paper.


\section*{Data availability}
The data underlying this article will be shared at request to the corresponding author.











\bsp	
\label{lastpage}
\end{document}